\documentclass[conference]{IEEEtran}

\usepackage{graphicx}  
\usepackage{epstopdf}

\usepackage{subfigure} 
\begin{document}

\title{Hierarchical Small-Worlds in Software Architecture}

\author{\authorblockN{Sergi Valverde\authorrefmark{1}}
\authorblockA{\authorrefmark{1}Complex Systems Lab\\
ICREA-UPF, Dr. Aiguader 88\\
08003 Barcelona, Spain\\
Email: svalverde@imim.es}
\and
\authorblockN{Ricard V. Sol\'e\authorrefmark{1}\authorrefmark{2}}
\authorblockA{\authorrefmark{2} Santa Fe Institute \\ 
1399 Hyde Park Road\\
Santa Fe, NM 87501, USA\\
Email: ricard.sole@upf.edu}
}


%


\maketitle

\begin{abstract}
In this paper, we present a complex network approach to the study of
software engineering.  We have found universal network patterns in a large collection of
 object-oriented (OO) software systems written in C++ and Java.  All the systems analyzed 
here display the small-world behavior, that is, the average distance between any 
  pair of classes is very small even when coupling is low and cohesion is high. 
 In addition, the structure of OO software is a very heterogeneous network
characterized by a degree distribution following a power-law with similar exponents. 
  We have investigated the origin of these universal patterns. Our study suggest that some features of OO programing languages, like encapsulation,  seem to be largely responsible for the small-world behavior.
 On the other hand, software heterogeneity is largely independent of the purpose and objectives of 
the particular system under study and appears to be related to a 
pattern of constrained growth. A number of software engineering topics 
may benefit from the present approach, including empirical software 
measurement and program comprehension.
\end{abstract}


%
\IEEEpeerreviewmaketitle

\section{Introduction}
It is over fifteen years since Norman Fenton outlined the need for 
a scientific basis of software measurement \cite{Fenton1994}. Such a theory 
is a prerequisite for any useful quantitative approach to software 
engineering, although little attention has been received from 
both practitioners and researchers.  Measurement is the process that assigns 
numbers or symbols to attributes of real-world entities.  
Unfortunately, many empirical studies of software measurements lack a forecast 
system that combines measurements and parameters in order to make 
quantitative predictions \cite{FentonBook}.  How we can overcome these
limitations?

Here we present a new approach to software engineering based on recent 
advances in complex networks \cite{Dorogovtsev, VitoReview, Bornholdt}. 
We study graph abstractions of software designs, where nodes 
represent software entities (i.e., classes and/or methods) and edges represent
 static relationships between  them (i.e., inheritance and association). We measure
graph attributes of software designs in order to find universal patterns of software
organization. Graph measures are not anew to software \cite{FentonBook}.   Empirical software
 studies assume there is a correlation between software design measures (i.e., lines of code, coupling, 
cohesion, modularity) and external features (like software reliability or development effort). Although 
good agreement has been observed in some cases, it is difficult to know if empirical mappings hold in
  general or not without appropriate models  \cite{Fenton1994}.  We have found that some graph 
measurements of software structures are (statistically) predictable. Moreover, they are within a definite 
range of values. Intriguingly, these patterns are almost independent of  functionality and  other 
external features. It seems that strong constraints limit the set of possible patterns that software 
structures can display.  These constraints might be useful to define useful reference models that
enable predictive software development processes.

\begin{figure}
\centering
\includegraphics[width=3.25in]{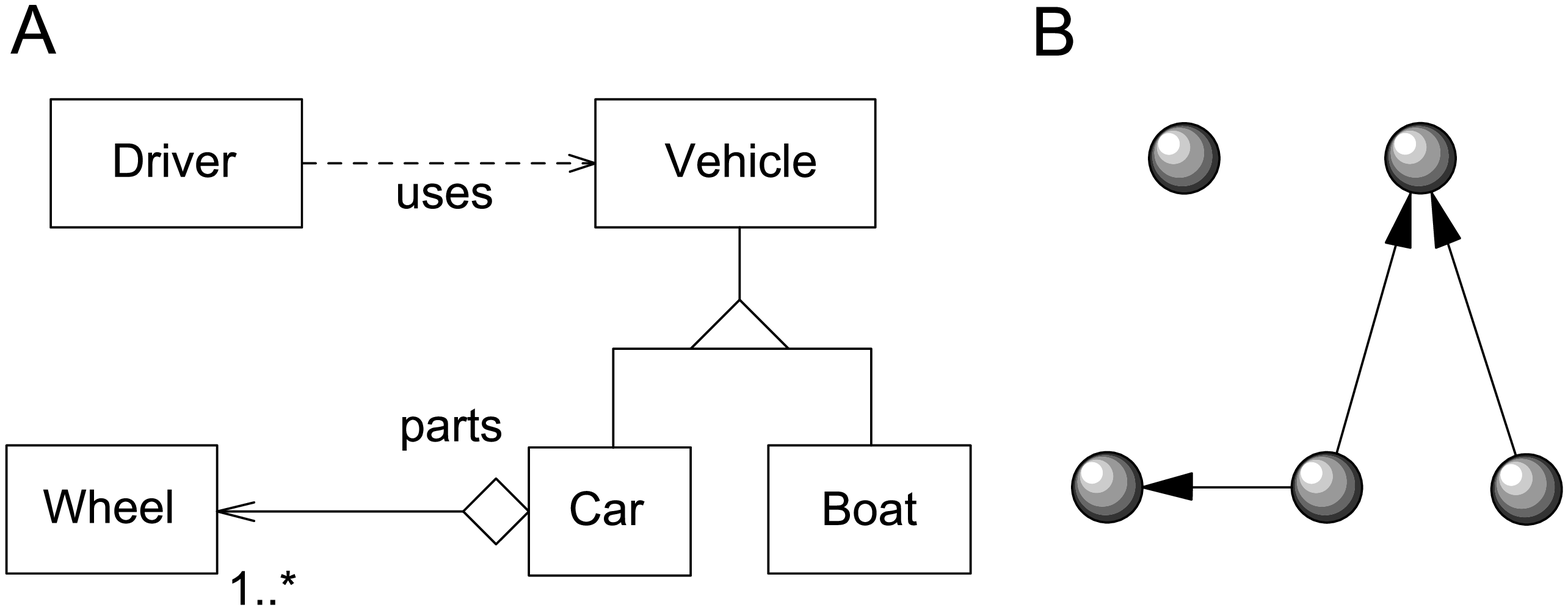}
\caption{(A) A simple UML class graph. (B) And its equivalent class graph (see text).
Every class maps to a single node in the class graph. Links in the class graph
represent member and inheritance relationships. Dynamic relationships 
 are not represented in the class graph (i.e, the "uses" relationship depicted with discontinuous link).
 We consider both the directed (B) and undirected (not shown) versions of a class graph.
}
\label{ClassGraph}
\end{figure}

 Object-oriented (OO) software systems display small-world (SW) behavior. Many
 real systems, including the WWW, food webs, and cellular networks \cite{Watts} are
 small-worlds, that is, they have high-degree of clustering and a small average distance.
Another common property of OO software is that probability distributions of structural
 attributes tend to follow skewed distributions with long tails \cite{Shepperd}. Heterogeneous
 metrics have been interpreted as an accident or the signal of rare, atypical behavior.  
 In this context, software researchers often avoid heterogeneity by
  manipulating the original distribution. Unfortunately,  this transformation hides important
   structural information and the true nature of OO software. Here, we show that the probability
   of a class to participate in $k$ relationships follows a scaling-law, that is, software
   designs are scale-free (SF) networks. We have validated the SW and SF behavior of
    OO software in many real systems, and thus suggesting they are universal
    features of software designs. In this paper, we explore the origin of these patterns. 
    Eventually,  we provide some tentative explanations but clearly more work is needed
    in this direction. 

 The regularities found here suggest that concepts and theories developed
by complex networks studies are useful in other
software engineering contexts, like program comprehension. For instance, 
OO software and the WWW share many structural features. Recent analyses of web
 graphs have  show the existence of some key pages
 called {\em hubs} and {\em authorities}\cite{Kleinberg99}.  Hubs are web pages having a
large number of links, like web directories or lists of personal pages. 
Authorities are pages that contain useful information and thus 
are pointed to by hubs. OO systems display a similar pattern, where a few (hub) classes 
have a large number of relationships.  Hub classes are excellent starting points for the program 
comprehension process.  A node centrality index might enable us to locate
key software components very quickly in a very large source code database 
(i.e, pageranks \cite{pagerank}).  In addition, we study a particular software
system and suggest that we can obtain useful information by comparing
different network representations of the same software system.
 
This paper is structured as follows. Section II defines class graphs, an abstraction that 
captures static structural features of object-oriented systems.  These
class graphs display universal features: they are small-worlds and scale-free
networks. Section III investigates the intrinsic origin of small-world behavior 
in class graph,  which seems to be related to the bipartite association between
 methods and classes.  Section IV proposes that class graphs are scale-free
  because they evolve under constraints and thus claiming for an external cause.
Finally, section V concludes the paper and outlines additional implications of the
network patterns found here in empirical software engineering and distributed
 software development.

\section{Class Graphs}

A \emph{class graph} (a software network) is a digraph $D=(V,L)$
that consists of the set $V$ of classes and the set of relationships 
$L = P \cup S$. There are two types of relationships:  a membership
 relationship $P=\{(v_i, v_j)\}$, i.e., read "$v_i$ has part $v_j$"; 
  and a reflexive and transitive relationship $S=\{(v_i, v_j)\}$, i.e.,  read 
  "$v_i$ is a subclass of $v_j$". However, and from now on, we will not make any
   distinction between these two relationships $P$ and $S$ and we only consider
    the full set of links $L$.   We discard any dynamic class relationship 
    from the graph definition.   For instance,  method invocation 
    (i.e., uses  relationship in fig. \ref{ClassGraph}A) is not 
represented in the class graph (compare with fig.\ref{ClassGraph}B). 
 Instead, we conceive nodes and links as black boxes hiding internal
  complexities that do not change the global structure. This bare-bones characterization
   enables us to detect global patterns in the static software structure. Ultimately, we hope that the analysis of class graphs will provide important  insights into high-level processes 
of software evolution. We also define  the undirected class graph (or undirected 
software network)   $G =(V, E)$ where  $E=\{\{v_i, v_j\}| (v_i, v_j) \in L \vee (v_j, v_i) \in L\}$ is the 
set of edges (see fig.\ref{ClassGraph}).


\begin{figure}
\centering
\includegraphics[width=3.5in]{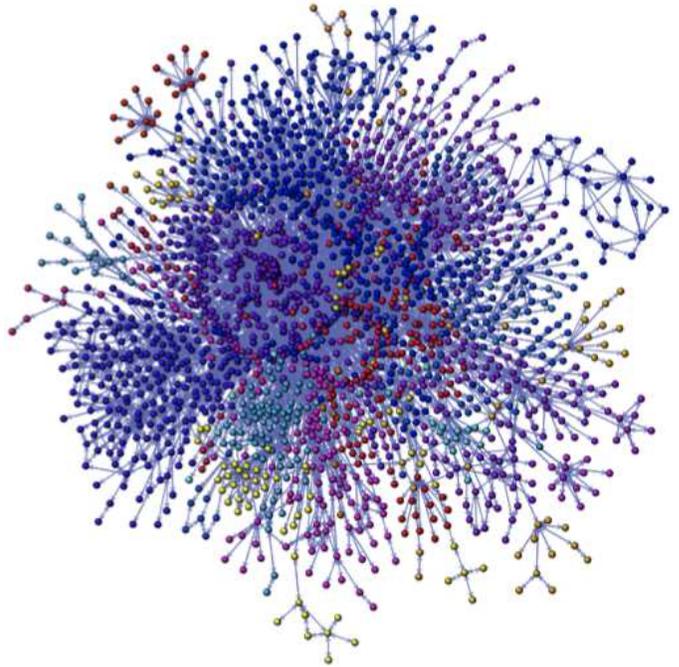}
\caption{Directed class graph for the computer game ProRally 2002. This scale-free and small-world 
network has $N = 1993$ classes. There is a clear modular organization and nodes naturally cluster
in different subsystems, i.e., 3d rendering, physical simulation, artificial intelligence, etc. Node color 
indicates the node subsystem.} 
\label{ProRally2002}
\end{figure}

Class graphs represent an important information space of
OO software systems. A prerequisite for software evolution and maintenance is that
software engineers recognize and understand the function 
performed in software.  This problem is aggravated in large software systems, 
where  source code navigation can turn easily into a bottleneck. 
The efficiency of program comprehension depends
on {\em general} and {\em new} knowledge \cite{Mayrhauser}. 
General knowledge  is independent of the particular software
 application. 
On the other hand,  new knowledge includes all the specific concepts and ideas
 regarding the particular software application. This includes  knowledge encoded 
in source code, which typically comprises several levels of abstraction.  
Each level of abstraction defines an {\em information space} or
subsets of the global information space representing the whole software 
system\cite{Robitaille}. These information spaces display an internal
structure that is navigated by software engineers to obtain
new knowledge and achieve {\em program comprehension}.  \cite{Sim99} 
further decomposes information retrieval in two different 
strategies: {\em Browsing} and {\em Searching}. Browsing is an exploration 
of high-level software entities while searching aims to low-level 
entities.    Efficient browsing requires an adequate software structure. For instance, modular
 software (i.e., a system that has been subdivided in disjoint chunks or modules with 
 clear boundaries) 
 enhances program comprehension 
  and minimizes the impact of changes.  In this context, we think that
 structural analyses of class  graphs might be useful to assess
  the performance of browsing and program comprehension in general.
   
\subsection{Data and Methods}
 
We have collected a large sample of 80 different software systems 
written in Java and C++.  This dataset represents a wide variety of
different software applications and it is large enough to be statistically
significant. We have recovered class graphs according to the 
definition given in the previous section. Actually, five systems provide full 
UML class diagrams: ProRally 2002 (a proprietary C++ videogame), 
Striker (C++ videogame),  JDK-A and JDK-B  (two largest connected 
components of Java 1.5) and Mudsi (a distributed JAVA application). 
These UML class diagrams are design documents released by their 
respective software developers. The mapping from UML class diagrams to class graphs
 is straightforward (see fig.\ref{ClassGraph}). The remaining software systems represent
 a diverse repertoire of Open Source (OS) applications written in C++. These 
class graphs were reverse-engineered with a simple lexical analysis of the 
C++/Java source codes (see Appendix I). Fig. .\ref{SoftLevel}e is the class graph for
the C++  code in fig. \ref{SoftLevel}a. In Table I there is a summary of 
graph measurements in a subset of software systems. Figure \ref{ProRally2002}
shows the class graph for the C++ videogame ProRally 2002 (see below).

\subsection{Connectance and Linear Growth}

\begin{figure}
\centering
\includegraphics[width=3.20in]{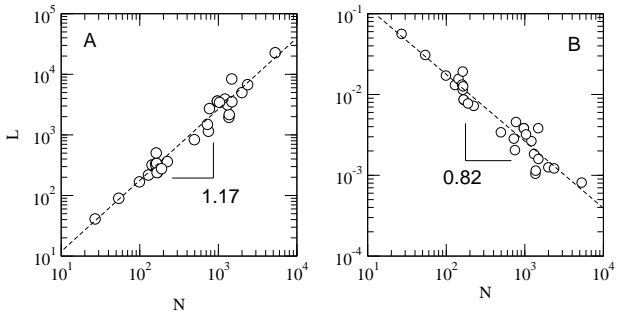}
\caption{Log-log plots of different size relationships in class graphs. Every point represents
a single class graph. (A) The number of links $L$ in class graphs scales linearly with number of
classes.  (B) The connectance or 
abundance of possible links $L/N(N-1)$ decays as the number of classes $N$ increases. 
The observed connectance is very small (about 0.1\% of all possible links).}
\label{LinkConnectivity}
\end{figure}

The number of links L=$|L|$ scales with the number of classes  $N=|C|$
in an almost linear way (see fig.\ref{LinkConnectivity}A):

\begin{equation}
\label{linksize}
L \sim N^{1.17} 
\end{equation}

This shows that class graphs are very sparse. In addition, this linear dependence 
between links and nodes means that every new class attaches  (on average)
 to an approximately constant number of existing classes. This fits very well
  the assumption of the linear growth in software systems \cite{Turski}. 
Define the {\em richness connectance} of a graph as the fraction 
of used links $L$ compared to the number $N(N-1)$ of links in the complete graph 
(self-referencing is avoided).  If $L$ scales 
linearly with $N$, then richness connectance will decay approximately 
as $1/N$ (see fig. \ref{LinkConnectivity}B).  Linear growth does not allow for
 extensive changes  to the large-scale class graph structure. Connectance decays 
 very fast and network size quickly saturates to a constant value. This saturation
  has been associated to a pattern of increasing complexity in software
   development \cite{Lehman}. 

\subsection{Class Graphs are Small-Worlds}

Watts and Strogatz found that many real networks display  short 
average path length and high clustering (or nonnegligible cliquishness)
\cite{Watts}. A network displaying these properties is called a 
{\emph small world} (SW). Given a  node $v_i$ with degree $k_i$ (i.e., the number 
of links attached to the node),  we define node clustering $C_i$ as the fraction of  actual 
number of triangles $t_i$ where the node $v_i$  participates in:

\begin{equation}
\label{localcluster}
C_i  = \frac{{t_i }}{{k_i (k_i  - 1)}}
\end{equation}

The clustering coefficient $C$ of a graph measures the proportion of triangles in the graph:

\begin{equation}
\label{clustercoeff}
C = \left\langle {\frac{2}{{k_i (k_i  - 1)}}\sum\limits_{j = 1}^N {A_{i,j} \left[ {\sum\limits_{k = 1}^N {A_{j,k} } } \right]} } \right\rangle 
\end{equation}

where $A$ is the adjacency matrix for the graph with $A_{i,j} =
1$ if node $v_i$ and $v_j$ are connected and $A_{i,j} = 0$ otherwise.
For random graphs, the clustering coefficient is inversely proportional to the graph size: 

\begin{equation}
\label{ranclustercoeff}
C_{rand}  \approx \frac{{ \left < k \right > }}{N}
\end{equation}

The clustering coefficient of a SW is significantly larger than 
the expected clustering  coefficient for the random network, $C >> C_{rand}$. 
Nodes in the SW are densely connected with its immediate neighborhood.

Average path length $d$ is a measure of the global connectivity, or the mean distance
 $d_{ij}$ required to navigate between any pair of nodes $v_i$ and $v_j$:

\begin{equation}
\label{pathlen}
d = \frac{1}{N}\sum\limits_{\forall i,j} {d_{ij} } 
\end{equation}

The average path length in random graphs is proportional to the logarithm of their size:

\begin{equation}
\label{ranpathlen}
d_{rand}  \approx \frac{{\log N}}{{\log \left < k \right > }}
\end{equation}

The average path length of a SW is as small as in the unrestricted random case 
$d \approx d_{rand}$, due to a few long-range edges (shortcuts) 
connecting distant regions of the network. Then, small average path 
length is compatible with a broad range of clustering coefficient 
values \cite{Watts}. This is a measure of network spread or compactness
that has been observed in different contexts, from the Internet to the social
 networks.  In these systems, it is useful to keep $d$ as low as possible. 
For example,  shortest paths often enable faster communications.   
On  the other hand, coupled oscillator systems with short average path 
lengths synchronize much faster than systems displaying 
longer paths \cite{Sync, Jinhu}.


\begin{table}[t]
\caption{Graph Measurements}
\center
\vspace{0.2cm}
\begin{tabular}{|c|c|c|c|c|c|c|}
\hline
\hline
Dataset & $N$ & $L$ & $d$ & $d_{rand}$ & $C$ & $C_{rand}$ \\
\hline
Mudsi & 168 & 241 & 2,88 & 4,95 & 0,244 & 0,017 \\
JDK-B & 1364 & 1947 & 5,97 & 6,80 & 0,225 & 0,002 \\
JDK-A & 1376 & 2162 & 5,40 & 6,28 & 0,159 & 0,002 \\
Prorally & 1993 & 4987 & 4,85 & 4,71 & 0,211 & 0,003 \\
Striker & 2356 & 6748 & 5,90 & 4,46 & 0,282 & 0,002 \\
gchempaint & 27 & 41 & 2,85 & 3,26 & 0,204 & 0,102 \\
4yp & 54 & 90 & 3,28 & 3,44 & 0,069 & 0,059 \\
Prospectus & 99 & 168 & 3,80 & 3,77 & 0,14 & 0,034 \\
eMule & 129 & 218 & 3,87 & 4,16 & 0,237 & 0,025 \\
Aime & 143 & 319 & 2,66 & 3,34 & 0,413 & 0,031 \\
Openvrml & 159 & 335 & 3,53 & 3,53 & 0,08 & 0,026 \\
gpdf & 162 & 300 & 4,02 & 3,93 & 0,303 & 0,022 \\
Dm & 162 & 254 & 4,32 & 4,45 & 0,304 & 0,019 \\
Bochs & 164 & 339 & 3,15 & 3,60 & 0,335 & 0,025 \\
Quanta & 166 & 239 & 4,31 & 5,03 & 0,198 & 0,017 \\
Fresco & 189 & 277 & 4,73 & 4,89 & 0,228 & 0,015 \\
Freetype & 224 & 363 & 4,29 & 4,71 & 0,193 & 0,014 \\
Yahoopops & 373 & 711 & 5,57 & 4,47 & 0,336 & 0,01 \\
Blender & 495 & 834 & 6,54 & 5,14 & 0,155 & 0,007 \\
GTK & 748 & 1147 & 5,87 & 5,91 & 0,081 & 0,004 \\
OIV & 1214 & 3903 & 3,99 & 3,82 & 0,122 & 0,005 \\
wxWindows & 1309 & 3144 & 4,03 & 4,62 & 0,235 & 0,004 \\
CS & 1488 & 3526 & 3,92 & 4,74 & 0,135 &  0,003 \\
\hline
\hline
\end{tabular}   
\end{table}

\begin{figure}
\centering
\includegraphics[width=3.35in]{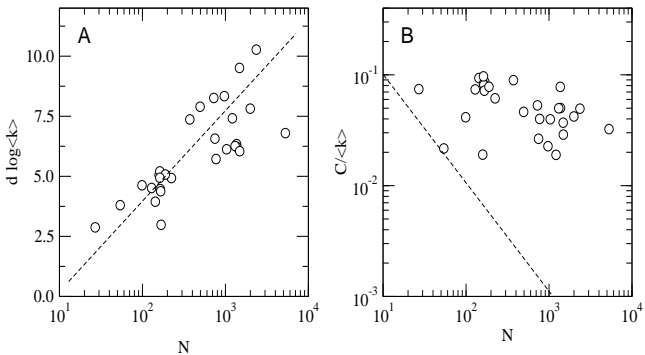}
\caption{Log-log plots validating the small-world model of class graphs. 
Every point corresponds to a single class graph. (A) Average path length vs class graph size. 
Normalized path length grows with the logarithm of  
number of classes, as expected in small world networks (see text). (B) 
Normalized clustering for the systems analysed here strongly departs from 
the predicted scaling relation followed by random graphs (dashed line). 
Class graphs are much more clustered (by orders of magnitude) than 
their random counterparts.}
\label{SWSoft}
\end{figure}

We have measured $d$ and $C$ in all the class graphs described above. Comparison 
with random predictions shows that class graphs are instances 
of small-worlds (see table I).  For every class graph, we have observed that
$C >> C_{rand}$ and $d \approx d_{rand}$.  In fig.\ref{SWSoft} we can clearly
 appreciate this  result: the clustering coefficient in class graphs is well above the
 random expectation while the average path length is rather small.  Actually, the 
values of $C$ seem rather independent from system size $N$. 
This is a common feature in hierarchical networks \cite{Oltzvai}.

\subsection{Small-World and Breakdown of Modularity}

Software systems are constantly evolving. A goal of software engineers is to minimize
the cost of software evolution by limiting the consequences of changes.  Some designs are 
better than others in this regard because they allow software engineers to make small changes 
without propagating many secondary changes to other software components.   In this context,
it would be desirable to have reliable estimates of future change costs. Unfortunately, we
still do not understand very well the properties of software development and maintenance.
Here, we propose that software maintenance is a global process and thus, it is very
difficult to predict the spreading of software changes.

\begin{figure}
\centering
\includegraphics[width=3.20in]{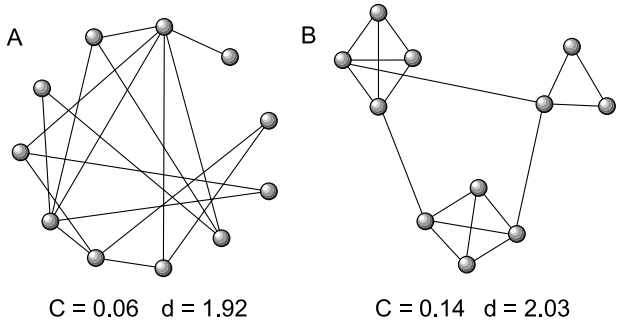}
\caption{Two different graphs with the same number of nodes $N=11$ and links $L=18$.
(A) Random graph with $<k> = L/N \approx 1.63$. This graph has small
average path length $d = d_{rand} = 1.92$ and low clustering $C = C_{rand} = 0.06$ 
(B) Modular graph with the same average degree. This graph is an small-world
because its average path length is comparable to random prediction
$d = 2.0 \approx d_{rand}$ and its clustering coefficient is about 
twice $C = 0.14 \approx 2 C_{rand}$. }
\label{SmallWorld}
\end{figure}

For example, compare the modular graph in fig.\ref{SmallWorld}B and 
the random graph in fig.\ref{SmallWorld}A. The graph in fig. fig.\ref{SmallWorld}B
 displays three, highly clustered, modules (i.e,  a module is a subset of nodes that exchange
many more links among them than with the rest of the network).  Here, modules
are interconnected to other modules by a single link and thus suggesting that 
internal changes in a module cannot affect other modules.  On the other hand, 
fig.\ref{SmallWorld}A is an example of highly coupled, loosely modular architecture. 
This is a random graph and all nodes belong to the same module. Changes in the 
random graph (fig.\ref{SmallWorld}A)  are more likely to affect many more nodes
 than changes in the modular graph (fig.\ref{SmallWorld}B).
Notice that local measures cannot separate random and modular
structures (i.e., the random graph and the modular graph have the same
average degree $\left < k \right >$). This suggests why empirical 
studies do not report significant correlations between local software measures 
  and change impact \cite{Kabaili}.  The state of a class relies on the state of all 
the other classes it references, including these classes referenced through a 
chain of intermmediate classes. 

There is ample evidence that many software projects have a natural 
tendency to become disordered structures \cite{Refactor}. This code degradation is often 
associated with a {\em breakdown of modularity}  that happens when changes are widely dispersed 
and affect many unrelated classes in apparently distant modules\cite{Decay}.  
We suggest that such breakdown of modularity might be related to
the emergence of the small-world behavior.  Recall that a highly-clustered class graph 
(i.e., a modular graph) becomes a small-world by the addition of a few shortcut links 
between dissimilar nodes ( i.e., a relationship between unrelated classes in different 
software modules).  Once the system displays small-world behavior, its average path 
length gets near the minimal value $d_{rand}$ and the software project might be closer
to a breakdown of modularity. In this context, we propose that software
engineers evaluate the risk of code degradation by measuring any significant
deviation of average path length (\ref{pathlen}).  This global measure could
be a better indicator of code degradation because it takes into account 
indirect effects. 

\subsection{Class Graphs are Scale-Free Networks}

\begin{figure}
\centering
\includegraphics[width=2.50in]{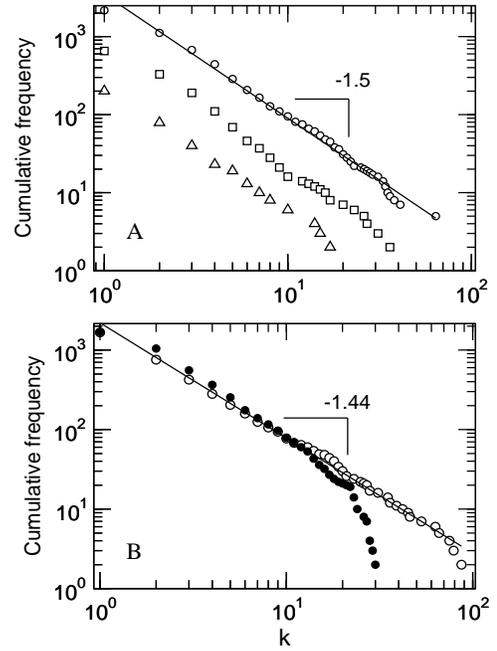}
\caption{(Top) Cumulative degree distributions for several class graphs:
eMule (N=129, triangles), Blender (N=495, squares) and CS (N=1488,
circles). All distributions have an exponent about -2.5 in spite of the obvious
differences in size and functionality. (Bottom) Asymmetry of in-degree (open
circles) and out-degree (filled circles) distributions for ProRally 2002. The
in-degree distribution is the probability that a given component is reused by
$k_{in}$ other components. Conversely, the out-degree distribution is the
probability that a component uses $k_{out}$ other components. 
Notice how outdegree distribution shows a sharp cutoff.}
\label{PowerLaws}
\end{figure}

Class graphs are highly heterogenous networks, where a very 
few classes participate in many relationships and the majority of classes
have one or two relationships \cite{Valverde2002}.  Highly connected classes are key
software components that keep the whole software system as a coherent
entity.  In this context, software designs are remarkably similar to many
other complex networks, like the WWW, the Internet and many biological
networks \cite{Dorogovtsev}. They are all examples of scale-free (SF) networks, that is,
they have a degree distribution that follows a scaling law, $P(k) \sim k^{-\gamma}$. 
As shown in figure \ref{PowerLaws} and in table II, class graphs are nice
instances of scale-free (SF) networks. The fact that all the graphs analysed here 
display SF structure, in spite of the obvious differences in size, functionality 
and other features, is an indication that strong constraints are at work 
in software evolution.  However, and contrary to the small-world feature
of class graphs, we suggest this scale-free behavior has an exogenous origin
(see below).

The cumulative degree distribution $P_> (k)$ reduces noise levels during
 the estimation of the scaling exponent  $\gamma$, 

\begin{equation}
\label{cumulative}
P_> (k) = \sum\limits_{k' > k} {P(k')} 
\end{equation}

If $P(k) \approx k^{-\gamma}$ then we have
$P_ >  (k) \approx \int {P(k')dk'} \approx k^{-\gamma+1}$. The
exponent $\gamma$ is estimated by linear regression in the
log-log plot (see figure 3b and figure 9).  For class graphs analyzed
here, we obtain $\gamma \approx 2.5$. On the other hand, in-degree and out-degree 
distributions of directed class graphs also follow power-laws,  $P_{in}(k) \sim k^{-\gamma_{in}}$ 
and $P_{out}(k) \sim k^{-\gamma_{out}}$.  Directed degree distributions display different
exponents from the undirected version.  Typically, we observe  $\gamma_{in} < \gamma$ 
and $\gamma_{out} > \gamma$.  In other words, if we look at the number of outgoing 
and  incoming links, the resulting degree distributions are different.  The in-degree distribution 
has a clear power-law tail while the out-degree distribution decays much faster.
A similar pattern has been observed in the web graph \cite{Albert99}. An extensive study of 
 the entire WWW in October 1999 used the webcrawl from Altavista to obtain empirical in-degree 
and out-degree distributions for a subset of the full web graph\cite{Broder}. They have shown 
that in- and out-degree distributions of the web graph are fitted by
scaling laws with exponents $\gamma_{in} = 2.1$ and  $\gamma_{out}= 2.7$. 
These exponents are very close to the average in-degree exponent $ \left <\gamma_{in} \right> = 2.2$ 
and the average out-degree exponent $\left <\gamma_{out} \right> = 2.8$ taken over all class graphs 
in table II.


%
\begin{table}[t]
\caption{Exponents of Cumulative Degree Distributions}
\center
\vspace{0.2cm}
\begin{tabular}{|c|c|c|c|}
\hline
\hline
Dataset & Degree & In-degree & Out-degree \\
\hline
Mudsi & 1.74 $\pm$ 0.04 & 1.20 $\pm$ 0.08 & 2.00 $\pm$ 0.05 \\
JDK-B & 1.55 $\pm$ 0.08 & 1.39 $\pm$ 0.05 & 2.30 $\pm$ 0.14 \\
JDK-A & 1.41 $\pm$ 0.02 & 1.18 $\pm$ 0.02 & 2.39 $\pm$ 0.14 \\
Prorally & 1.72 $\pm$ 0.03 & 1.44 $\pm$ 0.02 & 1.88 $\pm$ 0.10 \\
Striker & 1.70 $\pm$ 0.04 & 1.54 $\pm$ 0.03 & 1.73 $\pm$ 0.06 \\
gchempaint & 1.63 $\pm$ 0.31 & 1.11 $\pm$ 0.35 & 1.41 $\pm$ 0.12 \\
4yp & 1.54 $\pm$ 0.09 & 1.30 $\pm$ 0.05 & 1.59 $\pm$ 0.18 \\
Prospectus & 1.67 $\pm$ 0.09 & 1.13 $\pm$ 0.09 & 1.92 $\pm$ 0.27 \\
eMule & 1.58 $\pm$ 0.03 & 1.51 $\pm$ 0.07 & 1.42 $\pm$ 0.08 \\
Aime & 1.43 $\pm$ 0.05 & 1.30 $\pm$ 0.04 & 1.48 $\pm$ 0.07 \\
Openvrml & 1.34 $\pm$ 0.06 & 0.94 $\pm$ 0.05 & 1.59 $\pm$ 0.23 \\
gpdf & 1.64 $\pm$ 0.11 & 1.23 $\pm$ 0.10 & 1.76 $\pm$ 0.17 \\
Bochs & 1.37 $\pm$ 0.08 & 1.17 $\pm$ 0.09 &1.64 $\pm$ 0.20 \\
Quanta & 1.69 $\pm$ 0.10 & 1.55 $\pm$ 0.13 & 1.87 $\pm$ 0.13 \\
Fresco & 1.66 $\pm$ 0.09 & 1.14 $\pm$ 0.10 & 1.76 $\pm$ 0.19 \\
Freetype & 1.65 $\pm$ 0.07 & 1.42 $\pm$ 0.04 & 1.82 $\pm$ 0.16 \\
Yahoopops & 1.67 $\pm$ 0.05 & 1.46 $\pm$ 0.06 & 1.69 $\pm$ 0.05 \\
Blender & 1.64 $\pm$ 0.04 & 1.36 $\pm$ 0.05 & 2.04 $\pm$ 0.09 \\
GTK & 1.51 $\pm$ 0.04 & 1.22 $\pm$ 0.02 & 2.38 $\pm$ 0.20 \\
OIV & 1.43 $\pm$ 0.02 & 1.14 $\pm$ 0.03 & 2.10 $\pm$ 0.12 \\
wxWindows & 1.41 $\pm$ 0.03 & 1.11 $\pm$ 0.02 & 2.18 $\pm$ 0.12 \\
CS & 1.58 $\pm$ 0.02 & 1.22 $\pm$ 0.03 & 1.96 $\pm$ 0.09 \\
\hline
\hline
\end{tabular}   
\end{table}

\subsection{Related Work}

\begin{figure*}
\centering
\includegraphics[width=5.25in]{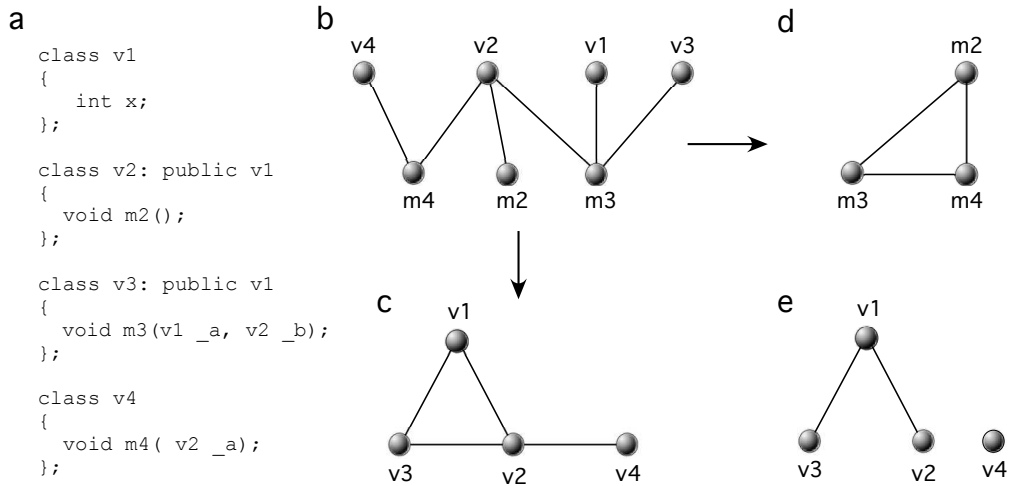}
\caption{(a) An example of C++ code and its: (b) bipartite association graph $B$, (c) its class projection $B_v$, (d) its method  projection $B_u$ and (e) its class graph $G$. }
\label{SoftLevel}
\end{figure*}
	
In order to measure software cohesion and coupling, 
\cite{Briand96} proposed to represent software designs with graphs, 
where nodes represent software entities and edges are relationships 
between entities. In this framework, a software module is a subset
of nodes (or subgraph) more densely connected 
than with the rest of the network. \cite{Briand96} explores 
what are the desirable properties of any cohesion and coupling measurement. 
At the coarsest level of description of a software system (or software architecture), 
a measure of coupling is the number of  edges exchanged between
 modules.  The cohesion of a module  comprising $k$  elements scales
  with the ratio $2t/k(k-1)$ where $t$ is the number of edges within
   the module. These coupling and cohesion definitions correlate with the 
number of edges $L$ and local clustering $C_i$ (\ref{localcluster}), 
respectively. In this context, a statistically valid model of class graph
 will provide useful estimators of coupling and cohesion in real systems.
 
Measurements are needed to assess the best design solution 
among different alternatives. From all the candidates in the 
solution space, we want to pick the one with the highest metric 
value \cite{Ince}.  It has been proposed that change impact 
defines one of the axes of this solution space. Some graph 
measurements from the logical structure of OO 
systems can be used as (static) estimators of change impact. 
A related definition is {\em alteration visibility} or the size 
of the set of classes affected by the change to a single 
class \cite{Chol}. A more detailed approach to impact analysis 
uses approximate algorithms to compute ripple effects 
from low-level source code features (see \cite{Ripple}\cite{Yau}). 
A distance measure very similar to path length was used
in \cite{Chol} in order to select the best choice when restructuring 
large software designs. However, \cite{Chol} proposed to 
measure path length between classes
belonging to a single module (i.e, intra-distance) while here we 
propose to measure path length between all the classes 
in the class graph (i.e, inter-distance). 

Chidamber and Kemerer (C\&K) presented a suite of object oriented 
metrics \cite{Chidamber} that seems to be related to our 
suite of graph measurements in class graphs . 
Several histograms of C\&K metrics display highly skewed distributions, i.e., 
 fig. 2 for the WMC (Weighted Methods per Class)
metric, fig. 14 for the CBO (Coupling between Object Classes) 
metric and fig. 16 for RFC (Response for a Class) metric 
in \cite {Chidamber}.  These histograms resemble power-laws like
the degree distribution of class graphs (see previous subsection). 
 Unfortunately, while C\&K metrics appear to be related to some of
 our graph measurements, no further comparison is possible because 
 no regression analysis of histograms is available from their study. 
Still, C\&K made a qualitative interpretation of extreme metric 
 values in terms  of "outliers" \cite{Chidamber}, which provides some
 evidence of heterogeneous software metrics.

In this context, there is a close relationship between {\em depth of inheritance tree} 
(or DIT, see \cite{Chidamber}) and degree of difficulty in 
understanding and comprehending the organization of 
object-oriented systems. Dvorak claimed that 
{\em "the deeper the level of the hierarchy, the greater the 
probability that a subclass will not consistently extend and/or 
specialize the concept of its superclass"} \cite{Dvorak},
that is, excessively deep class hierarchies are complex to develop. 
Evidence of positive correlation between DIT and the likelyhood
of faults in OO systems is given in \cite{Basili1996}.  
We should expected some correlation between DIT and average 
path length $d$ of class graphs because inheritance tree
is a subset of the class graph, suggesting how DIT might
be closely related to the small-world. Very low values of DIT found 
in the study of Li and Henry \cite{LiHenry} provide some empirical support
to this hypothesis. 

\section{Class-Method Association Graphs}

We have shown that class graphs are small-worlds. Here, we investigate the possibility that
 small-world can spontaneously emerge in an evolving OO software system.   We provide
  empirical and theoretical support to this conjecture by modeling the hierarchical structure 
  of OO software with a bipartite association between classes and methods.  In Java and C++,
we conceive a software system as a set of interrelated classes. These
classes are further decomposed into data members (or variables) and code methods (a
method is the OO equivalent of subroutines in Fortran, C or Pascal).  This 
hierarchical organization of OO software can be represented with a bipartite
association graph $B= (V, U, E)$ where $V=\{v_i\}$ is the set of classes and 
$U=\{m_i\}$ is  the set of methods and $E = \{\{v_i, m_i\}\}$ is the set of dependencies
between  classes and methods.  Also, $N=|V|$ is the number of classes 
and $M=|U|$ is the number of methods.  We have an edge $\{v_i, m_j\} \in E$ when
 class $v_i$ appears in the parameter list of method $m_j$. In addition, 
a class is always a parameter of its own collection of methods (i.e, {\em self}  or {\em this}
keyword).  We can recover this bipartite graph with a simple algorithm (see Appendix II). 
Figure \ref{SoftLevel} illustrates a small C++ code (see fig. \ref{SoftLevel}a) and its
 corresponding bipartite association graph (see fig. \ref{SoftLevel}b). 

 We define the (discrete) generating functions $\mu(n)$ and $\nu(n)$ for the bipartite 
 graph: 

\begin{equation}
\label{genfunmu}
\mu(n) = \sum\limits_k {k^n P_u } (k)
\end{equation}

and

\begin{equation}
\label{genfunnu}
\nu(n)  = \sum\limits_k {k^n P_v } (k)
\end{equation}

where $n = 1, 2,...$ and $P_u(k)$ is the fraction of $U$ nodes having $k$ edges 
and $P_v(k)$ is the fraction of $V$ nodes having $k$ edges. First moments 
$\mu = \mu(1)$ and $\nu = \nu(1)$ indicate the average method degree and the
 average class degree, respectively. It is easy to check that $M\nu = N\mu$. 

 The one-mode projection (or unipartite) network expresses  connections between
  nodes of the same kind (see fig. \ref{SoftLevel}c,d). We have two one-mode projections $B_v=(V,E_v)$ 
(i.e., so-called class projection) and $B_u = (U, E_u)$ (i.e., so-called method projection) from the
bipartite association method-class graph.  Formally, we define $A$ as the adjacency matrix of the 
bipartite network $B$, where  $A_{i,j} = 1$ if $\{v_i, u_i\} \in E$ and $A_{i,j}= 0$ otherwise. The adjacency matrix $A^V$ for the one-mode projection $B_v$ is related to the adjacency matrix $A$ by:

\begin{equation}
\label{projection1}
A_{i,j}^V  = \sum\limits_k {A_{ik} A_{jk} } 
\end{equation}

A similar relation holds between the adjacency matrix $A^U$ of
projection $B_u$ and the adjacency matrix $A$ of the bipartite network $B$:

\begin{equation}
\label{projection2}
A_{i,j}^U  = \sum\limits_k {A_{ki} A_{kj} } 
\end{equation}

Netwman et al. have shown that one-mode projections 
must be small-worlds even when the bipartite association is random\cite{Newman2000}. 
 Social networks display high-clustering coefficients because agents follow a natural
  tendency  to group together in communities. Moreover, the addition of a few shortcuts 
  between distant agents in clustered communities yields to small average path lengths. 
 Assuming that bipartite association $B$ is random, we have that the average path length
  between two classes in $B_v$ will be very small,

\begin{equation}
d(B_v ) = \frac{{\log N}}{{\log z}}
\end{equation}

and correspondingly for the method projection $B_u$ we have,

\begin{equation}
d(B_u ) = \frac{{\log M}}{{\log z}}
\end{equation}

where $z = \mu \nu$ is the expected average degree for the one-mode projection.  
The clustering coefficient for the one-mode projection $B_v$ will be very high:

\begin{equation}
C(B_v ) = \frac{1}{{\mu  + 1}}
\end{equation}

and for $B_u$, 

\begin{equation}
C(B_u ) = \frac{1}{{\nu  + 1}}
\end{equation}

Then, the above suggests that partitioning a OO software system into classes
and methods is very likely to result into a highly-clustered software structure with
 small average path lengths. Moreover, this seems to be largely independent of 
 the specific association between methods and classes. The clustering coefficient
 and mean path length only depend on the average connectivity
  of classes and methods.  The small-world behavior of class graphs does not
  appear to be an additional requirement selected by software engineers but
  an unavoidable feature associated to the hierarchical nature of OO software
  systems.
 
\subsection{Comparison between Class Graphs and Class Projections}

\begin{figure}
\centering
\includegraphics[width=3.5in]{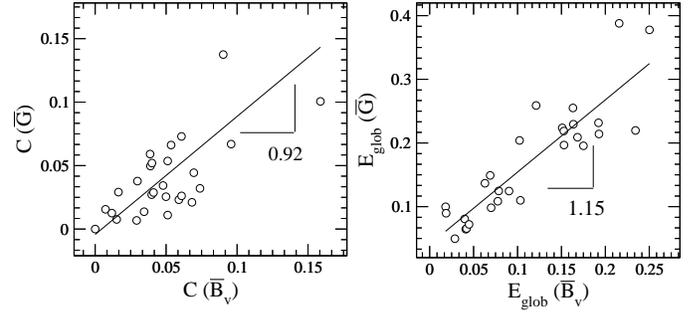}
\caption{Comparison between local and global measurements in class graphs and
class projections. Every point represents the pair of graphs ($\bar G$, $\bar B_v$)
from a single software system. (A) The clustering coefficient in class graphs 
scales linearly (fitting slope = $0.95 \approx 1$) with clustering coefficient in class projections. (B) 
Global efficiency in class graphs scales linearly (fitting slope = $1.15 \approx 1$)
with global efficiency in class projections. Good agreement suggests that
statistical features of class graphs, like the small-world property, may be inherited 
from class projections.}
\label{ClassProject}
\end{figure}

Beyond specific topological patterns displayed in any OS software
system we can investigate how well (in a statistical sense) the class projection
explains structural patterns displayed by the class graph. Here, we
are more interested in the structural properties of the {\em average} OS software 
system.  In this context, we have found very good agreement between
local and global measures of class graphs and class projections. In order to enable a
 meaningful comparison between the class projection $B_v$ and the class graph $G$,
  we must ensure they have the same number of nodes and links. First, we obtain the filtered class
  graph $\bar G =(\bar V, \bar E)$ by removing $G$ nodes without edges in $B_v=(\bar V, E_v)$.
    On the hand, the class  projection $B_v$ often displays more edges than the filtered
     class graph $\bar G$, that is, $|E_v| \ge |\bar E|$. Then, we remove 
     a fraction $p = 1- |\bar E|/|E_v|$  from the original class projection to obtain the
     filtered class projection $\bar B_v = (\bar V, \bar E_v)$ having the same number of
     nodes and edges in the filtered class graph $\bar G$.  For the systems
     analyzed here, the average edge removal probability is $\left < p \right > \approx 0.54$. 
     
Clustering coefficient of (filtered) class graphs scales almost linearly with clustering coefficient
 measured in the (filtered) class projections (see fig. \ref{ClassProject}A):  

\begin{equation}
C(\bar G) = 0.92 C(\bar B_v) \pm O(1)
\end{equation}

Edge removal can leave disconnected nodes in the class projection.  Average
path length $d$ cannot be computed in disconnected networks because $d_{i,j}=\infty$.
Fortunately, we can use the global efficiency $E_{glob}$ measure that is formally 
equivalent to average path length. Global efficiency of an undirected graph $G$ 
is defined as follows \cite{VitoReview}:

\begin{equation}
E_{glob} (G) = \frac{1}{{N(N - 1)}}\sum\limits_{i \ne j \in G} {\frac{1}{{d_{i,j} }}} 
\label{globaleffic}
\end{equation}

where $0 \le E_{glob} (G) \le 1$. Note that the maximum value of global
efficiency is attained when $G$ is the complete graph having $N(N-1)/2$
possible edges and the minimum value indicates that $G$ has no edges,
i.e., the graph is completely disconnected. We have found that global efficiency in
 class graphs scales with global efficiency of class projections:

\begin{equation}
E_{glob}(\bar G) = 1.15 E_{glob}(\bar B_v) \pm O(1)
\end {equation}

Good agreement of local and global measurements in the class graph $\bar G$ and 
the class projection $\bar B_v$ provides support that the SW behavior of
 class graphs is an invariant feature of {\em any }ÊOO software system. OO programming
 requires that related code and data cluster together in the same class, and thus resulting
  in high clustering coefficients. However, methods cross class boundaries when they use
  data (and methods) from other external classes. These eventual interactions among
   unrelated software entities yield small average path lengths.  
    
\subsection{Case Study: Stellarium}

\begin{figure}
\centering
\includegraphics[width=3.0in]{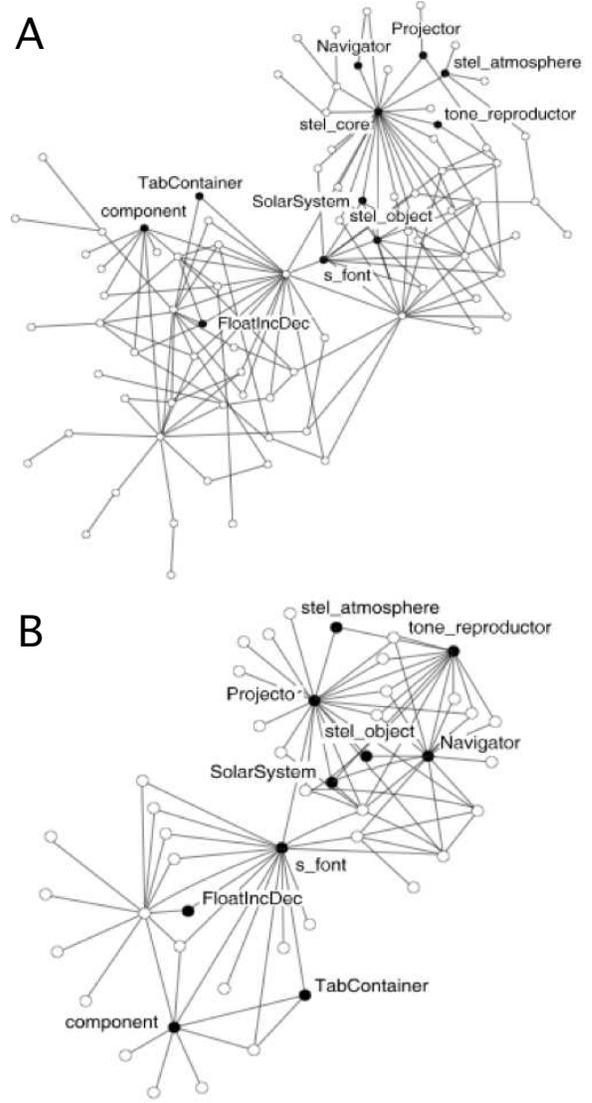}
\caption{Comparison between the (A) class graph $\bar G$ and the (B) class projection 
$\bar B_v$ for the OS software Stellarium. These graphs have the same 
size $N= 101$ and number of edges $L=162$. Here we display the largest
connected component of each graph, thus explaining the apparent size difference.
These networks have different adjacency matrices but identification of nodes 
(displayed with solid black circles) in $\bar G$ and in $\bar B_v$ suggests they
have the same modular structure. Notice the class {\tt\small s\_font} is at 
the boundary separating the two main modules.} 
\label{Modularity}
\end{figure}

We illustrate a detailed comparison between class graph and class projection (see above) with 
the OS software Stellarium (http://stellarium.free.fr).  This comparison suggests how useful 
is to analyze and compare several network representations during reverse engineering
and program comprehension. Stellarium is written in C++ and computes
 the position of stars and other space bodies in real-time.  Figure \ref{Modularity} shows the largest connected components of class graph $\bar G$ and 
class projection $\bar B_v$ recovered from the C++ source code our
reconstruction algorithm. Looking at the class 
graph (see fig.\ref{Modularity}A) we notice two well-defined communities or modules
 in Stellarium. Comparison between the class graphs and the class projection 
 indicates that modularity is preserved across multiple levels of the software hierarchy
 (see fig.\ref{Modularity}B).  Indeed, every software system analyzed here follows this pattern: 
 global features are  preserved across levels while individual nodes might play different roles
  depending on the network representation. Table III and table IV summarize 
  individual node measurements (classes are highlighted with solid circles in figure \ref{Modularity}).  

\begin{table}[t]
\caption{Node Measurements in $\bar G$}
\center
\begin{tabular}{|c|c|c|c|c|}
\hline
\hline

Node & $k_{in}$ & $k_{out}$ & $C$ & $BC(\times10^2)$ \\
\hline
stel\_core & 3  & 22 & 0.02 & 3 \\
Navigator & 1  & 1 & 1 & 0 \\
s\_font & 9  &  1 & 0.2 & 0.03 \\
Projector       & 3  & 0 & 0 & 0 \\
tone\_reproductor & 2 & 0 & 0 & 0 \\
stel\_atmosphere & 1 & 3 & 0 & 0.2 \\
SolarSystem & 1 & 3 & 0.66 & 0.01 \\
stel\_object & 5 & 1 & 0.26  & 0.03 \\
TabContainer & 1 & 1 &  1       & 0 \\
component & 6   & 2 & 0.03  & 1.2 \\
FloatIncDec & 1 & 3 & 0.33  & 0.06 \\
\hline
\hline
\end{tabular}   
\caption{Node Measurements in $\bar B_v$}
\center
\begin{tabular}{|c|c|c|c|c|}
\hline
\hline
Node & $k$ & $C$ \\
\hline
stel\_core      & 2 & 0 \\
Navigator & 15 & 0.21 \\
s\_font & 17 & 0.11 \\
Projector & 21 & 0.12 \\
tone\_reproductor & 13 & 0.24 \\
stel\_atmosphere & 2 & 1 \\
SolarSystem & 4 & 1 \\
stel\_object & 2 & 1 \\
TabContainer & 3 & 1 \\
component       & 8 & 0.21 \\
FloatIncDec & 2 & 1 \\
\hline
\hline
\end{tabular}   
\end{table}

For instance, Class {\tt Projector} belongs to the same module in $\bar G$ and 
in $\bar B_v$. However, {\tt Projector} is a hub in $\bar B_v$ but has few connections 
in $\bar G$ (see fig.\ref{Modularity}).  An opposite example 
is the class {\tt stel\_core}, which is a hub in $ \bar G$ but has only 
two connections in $\bar B_v$  (in fact, this node belongs to a 
disconnected subgraph not shown in the above figure).  
Class {\tt stel\_core} relies in many other classes ($k_{out} = 22$) and
is the main application dispatcher in Stellarium. {\tt stel\_core} is the starting point 
of many code reviews and thus, is frequently visited by Stellarium engineers. 
Consequently, this class displays the highest centrality value $BC= 0.03$, 
measured as the total number of shortest paths passing through a node 
\cite{VitoReview}. The second largest centrality value $BC=0.012$ is displayed 
by class {\tt component} representing the base class of any user interface 
control in Stellarium. 
On the other hand, class {\tt Projector} is an instance of the 
{\em state} design pattern \cite{DesignPatterns} and keeps the 
graphical application state (i.e., projection matrix, observer 
coordinates, etc.). {\tt Projector} is referenced by many methods, explaining 
why it has many connections ($k=21$) in $\bar B_v$.

\section{Scale-free and Evolution Constraints}

Scale-free networks can be obtained with simple generative models
of network evolution. These models  have two main components: network growth and
preferential attachment  \cite{Science1999}.   This suggests that scale-free nature
of OO software systems stems from the evolutionary process. However, the preferential 
attachment model fails to reproduce the structure of class graphs. For example, the predicted
 exponent $\gamma = 3$ for the model is different from the observed 
 exponent $\gamma \approx 2.5$. In addition,  the clustering coefficient for preferential
 attachment is very low and thus, cannot explain why class graphs are highly clustered. 

\begin{figure}
\centering
\includegraphics[width=3.2in]{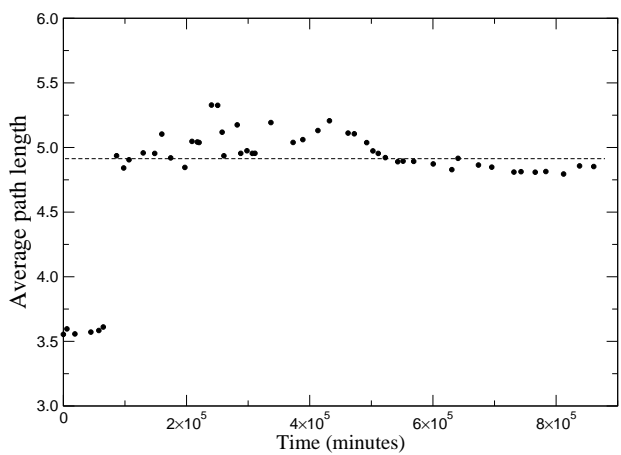}
\caption{Evolution of average path length during ProRally 2002 development.}
\label{prdist}
\end{figure}

 Following an approach similar to other empirical studies of software evolution \cite{Kemerer99},
  we have collected longitudinal data from the evolution of ProRally 2002, a large, priopietary
   computer game from Ubi Soft that was developed by 20 software engineers (see fig.\ref{ProRally2002}).
  We have recovered 176 intermediate class graphs comprising two years 
of development.  From this dataset, we have analyzed time
series of the number of nodes $N(t)$, number of links $L(t)$ and average path length $d(t)$. 
The growth pattern followed by ProRally 2002 was approximately linear in 
$N$ and $L$ (consistently with the general observation made in section IV.C)
and the final class graph has $N= 1993$ and $L=4987$ links (see fig. \ref{ProRally2002}).
Table I and II reports some network measurements for the final ProRally 2002 class graph. 
The time evolution of average path length $d$ in ProRally 2002 quickly saturates with
 $d \approx 5$ after a brief transient (see fig. \ref{prdist}).  This constant growth pattern
  in the time evolution  of ProRally 2002 yields an heterogeneous $P(k)$. As shown by Puniyani
and Lukose, growing random networks under the constraint of
constant diameter must display scale-free architecture and with 
a scaling exponent $\gamma \in [2, 3]$  \cite{Puniyani}. Specifically, they found that:

\begin{equation}
P(k) \approx k^{3 - \frac{\alpha }{\beta }} 
\end{equation}

where $\alpha \le 1$ is an exponent relating network size $N$ with 
degree fluctuations:

\begin{equation}
N^\alpha   = \frac{1}{{\left\langle k \right\rangle }}\int_k {k^2 P(k)dk} 
\end{equation}

and $\beta$ is an exponent linking the degree cutoff $k_c$ (see subsection II.D)
with network size $N$, i.e., 

\begin{equation}
k_c \approx N^\beta  
\end{equation}

Using our dataset, we estimate $\beta = 0.62 \pm 0.09$ and $\alpha = 0.42 \pm 0.08$.
This predicts the scaling exponent  $\gamma \approx 2.59$ to be compared with the
average exponent over all systems $\left\langle \gamma  \right\rangle  = 2.57 \pm 0.07$ (computed
from table II). The scaling law in the cutoff $k_c(N)$ allows us to provide an analytic 
calculation of the scaling between $L$ and $N$. The following integral gives the general 
relationship between $L$ and $N$:

\begin{equation}
L = N\int\limits_0^\infty  {kP(k)dk} 
\end{equation}

Here, we have,
\[
L \approx \frac{N}{{k_c^{2 - \gamma } }} \\ 
\]
\[
\approx N(N^\beta  )^{\gamma  - 2}  = N^{1 + \beta (\gamma  - 2)}  \approx N^{1.22}  \\ 
\]
in very good agreement with the exponent obtained in section II.E for class graphs.
Keeping the average path length constant during class graph evolution yields an 
heterogeneous degree distribution with the observed exponent. However, we were unable to
find any intrinsic explanation to this constraint. A possible explanation is an exogenous
pressure related to communication constraints in distributed software teams \cite{Conway}
(see below).

\section{Conclusion}

The structure and operation of software systems can be studied at
different levels of organization, from the small and simple (instructions) 
to the large and complex (the modular architecture).  The signature
of complex software organization is an heterogeneous and hierarchical
 network.  This pattern partly explains why it is difficult to find a clear,
 nice decomposition of software systems.  Moreover, the role of
 broad distributions in software measurements have been largely
 dismissed.  Researchers treat these distributions like normally
 distributed \cite{Shepperd}. Such transformation losses a significant
  amount of information and hides the true nature of software . Instead, we must
  address heterogeneous distributions with appropriate tools.  This knowledge
could be crucial to develop future software systems. For example, degree distributions
 predict how many classes have more than, say, a hundred, data members in a
  future class diagram of doubled size. 
 Notice that using non power-law expressions of degree distribution inevitably 
yield inaccurate predictions. In conclusion, we must abandon 
reductionistic descriptions of OO systems and replace this view by 
large-scale statistical characterizations preserving the structural variability. 

A more general question concerns the uselfuness of single-valued metrics. 
For example, the distribution of class sizes (measured as number 
of lines of code, NLOC) encodes more information than the integral 
value or system size. Given a NLOC value (say, 10KLOC),
there are many distributions satisfying this integral value. That is, 
NLOC is an ambiguous measure that provides less information than
the original size distribution.  We can compute average path length
 and the average clustering coefficient from probability distributions of 
 basic graph metrics, i. e. connectivity. There is an important source of
  information in the probability distributions of software measurements.

On the other hand, large-scale software development is a 
social task. Interaction between software engineers might have
 an influence in the organization and structuring of source code bases. 
 For example,  open-source developments are geographically
 distributed. These software teams face pressing communication 
 and coordination problems that require specific software structures 
according to the social organization: {\em "organizations which design systems are constrained 
to produce designs which are copies of the communications 
structure of these organizations"} (Conway's Law)\cite{Conway}. 
Under this social perspective, software is viewed simultaneously 
as the product and the vehicle that enables efficient communication 
between software engineers, i.e, the communication medium.
Separated software modules minimize communication overheads, which is
 the bottleneck in distributed software developments \cite{Herbsleb}.
In this context, it should be interesting to study how the patterns described
here relate to distributed software development.

Finally, while our study focused on static analyses of source code, 
recent studies on object graphs have revealed similar patterns in 
the distribution of run-time connections between objects in 
several programs \cite{Potanin}. This similarity suggests a link 
between structural and dynamical features of OO software that should
be investigated in future studies.


\appendices
\section{Class graph reconstruction algorithm}

The following algorithm reconstructs the class graph 
from a collection of Java/C++ header files (comments are 
highlighted in italics).  The class {\tt Digraph} implements a directed graph. 
The method {\tt Digraph::AddLink(c1, c2)} tests if class names 
{\tt c1} and {\tt c2} have been already inserted in the graph. If not, 
they are inserted correspondingly. We discard repeated links $(c_1, c_2)$.
 There is  distinction between public, private or protected
attributes. Finally, the algorithm outputs the directed (and undirected)
network versions to a file. 

{\tt\small 

\vspace{0.2in}

Digraph D; {\it // class graph $D = (C, L)$}

String c1, c2;  {\it // class names}

{\bf FOR} every header file {\bf DO}

\hspace{0.2in}{\bf WHILE} (not end of file) {\bf DO}

\hspace{0.4in}{\it // Find class declaration}

\hspace{0.4in}Look for 'class' keyword;

\hspace{0.4in}c1 = get\_class\_name();  {\it $// c_1 \in C$}

\hspace{0.4in}{\it // Test if inheritance relationship ("is a")}

\hspace{0.4in}{\bf IF} (next sequence is ': public') {\bf THEN}

\hspace{0.6in}c2 = get\_parent\_class(); {\it $// c_2 \in C$}

\hspace{0.6in}D.AddLink(c1, c2);  {\it $// (c_1 , c_2) \in  L$}

\hspace{0.4in}{\bf ENDIF}

\hspace{0.4in}{\it // get attributes ("has a")}

\hspace{0.4in}{\bf WHILE} (not end of class) {\bf DO}

\hspace{0.6in}Look for attribute declaration;

\hspace{0.6in}c2 = get\_attribute\_class(); {\it $// c_2 \in C$}

\hspace{0.6in}D.AddLink(c1, c2);  {\it $// (c_1 , c_2) \in  L$}

\hspace{0.4in}{\bf END}

\hspace{0.2in}{\bf END}

{\bf END}

D.Output();
}

\section{Association graph reconstruction algorithm }

The following algorithm recovers the bipartite association class-method graph 
from a collection of C++ header files.  The class {\tt Bipartite} implements 
a bipartite graph. There is a method {\tt Bipartite::AddLink(u, v)} that checks
if method {\tt u} and class {\tt v} have been already inserted in the graph.
We assume that methods have unique identifiers. Methods having the same name 
can still be differentiated because they belong to different classes 
(and classes cannot have the same name).

{\tt\small 

\vspace{0.2in}

Bipartite B; {\it // association graph $B = (V,U,E)$}

String v1, v2;  {\it // class names}

String u;  {\it // method name}

{\bf FOR} every header file {\bf DO}

\hspace{0.2in}{\bf WHILE} (not end of file) {\bf DO}

\hspace{0.4in}{\it // Find class declaration}

\hspace{0.4in}Look for 'class' keyword;

\hspace{0.4in}v1 = get\_class\_name();  {\it $// v_1 \in V$}

\hspace{0.4in}{\bf WHILE} (not end of class) {\bf DO}

\hspace{0.6in}Look for method declaration;

\hspace{0.6in}u = get\_method\_name(); {\it $// u\in U$}

\hspace{0.6in}D.AddLink(u, v1);  {\it $// \{u , v1\} \in  E$}

\hspace{0.6in}{\bf WHILE} (not end of method) {\bf DO}

\hspace{0.8in}v2 = get\_parameter\_class();  {\it $// v_2 \in V$}

\hspace{0.8in}D.AddLink(u, v2);  {\it $// \{u , v2\} \in  E$}

\hspace{0.6in}{\bf END}

\hspace{0.4in}{\bf END}

\hspace{0.2in}{\bf END}

{\bf END}

B.Output();
}


\section*{Acknowledgment}
The authors would like to thank Marcus Frean, Harold Fellerman, David Hales
 and Jos\'e M. Montoya for their careful comments and suggestions. S.V thanks 
 helpful  discussions with Alex Rodriguez, Jos\'e A. Andreu, Jos\'e Paredes,
Antoni Zamora, Luis Cascante, and Xavier Gir\'o from the ProRally 2002 team and
Fabien Ch\'ereau from the Stellarium team. This work has been 
supported by grants BFM2001-2154 and by the EU within the 6th 
Framework Program under contract 001907 (DELIS) and by the 
Santa Fe Institute.



%

\end{document}